\documentclass[final,5p,times,twocolumn]{elsarticle}
\usepackage{textgreek}
\usepackage{verbatim} 
\usepackage{amssymb}
\usepackage{url}
\usepackage{amsmath,amssymb}
\usepackage{mathtools} 
\usepackage[symbol]{footmisc}
\usepackage{subcaption}
\usepackage{capt-of}
\usepackage{bigints}
\usepackage{bm}
\usepackage{xcolor} 
\usepackage{soul} 
\usepackage{lipsum}
\def\be{\begin{eqnarray}}
\def\ee{\end{eqnarray}}
\def\nn{\nonumber}

\journal{Physics Letters B}
\makeatletter
\def\ps@pprintTitle{%
  \let\@oddhead\@empty
  \let\@evenhead\@empty
  \let\@oddfoot\@empty
  \let\@evenfoot\@oddfoot
}
\makeatother
\begin{document}
\begin{frontmatter}
\title{Gluon Wigner distributions with transverse polarization at non-zero skewness}
\author[first]{Sujit Jana}
\author[first]{Kenil Solanki}
\author[first]{Vikash Kumar Ojha\corref{cor1}}
\cortext[cor1]{Corresponding author}
\ead{vko@phy.svnit.ac.in}
\affiliation[first]{organization={Department of Physics},
            addressline={Sardar Vallabhbhai National Institute of Technology}, 
            city={Surat},
            postcode={395 007}, 
            state={Gujarat},
            country={India}}

 \begin{abstract}
We investigate the gluon Wigner distributions at non-zero skewness using light-front wave functions (LFWFs) in the dressed quark model, where the target state is a quark dressed with a gluon in the leading-order Fock space expansion. Our analysis focuses on the configurations in which the gluon, the target, or both are transversely polarized. We derive analytical expressions for the Wigner distributions in the boost-invariant longitudinal space ($\sigma$) for transversely polarized configurations and observe a diffraction-like oscillatory pattern in 
$\sigma$-space, analogous to that reported earlier for unpolarized and longitudinally polarized gluons.
 \end{abstract}
 \end{frontmatter}

 
\section{Introduction}
One of the fundamental objectives in quantum chromodynamics (QCD) is to understand the three-dimensional structure of hadrons through their underlying quark and gluon degrees of freedom \cite{Accardi:2012qut,anderle2021electron}. Among the most comprehensive tools for hadronic tomography are Wigner distributions, which encode the simultaneous information of partons in position and momentum space, akin to phase-space distributions in quantum mechanics \cite{Ji:2003ak,Belitsky:2003nz}. Wigner distributions are quasi-probability functions extensively used in quantum optics, signal processing, and quantum field theory to study the phase-space structure and dynamics of quantum systems \cite{bastiaans1979wigner, forbes2000wigner, radhakrishnan2022wigner}. While Wigner distributions are not directly observable due to their quantum nature and lack of probabilistic interpretation, they are rich in structural information and provide insight into phenomena such as orbital angular momentum, spin–orbit correlations, and partonic correlations within the nucleon \cite{lorce2011quark,Chakrabarti:2016yuw,Lorce:2011ni,mukherjee2014quark}.

Numerous studies have investigated quark Wigner distributions, taking into account both zero and non-zero skewness scenarios \cite{ojha2023quark, Luan:2024nwc,Broniowski:2023jgk, Han:2022tlh, Yang:2025neu}. In contrast, gluon Wigner distributions, especially those involving polarization effects and skewed kinematics, remain relatively less studied. Given the crucial role gluons play in the nucleon's momentum, spin, and small-$x$ dynamics, a deeper understanding of their phase-space distributions is essential. Probing the role of gluons inside the proton, particularly their contributions to the proton’s spin through helicity and orbital angular momentum is one of the primary objectives of the Electron–Ion Collider (EIC) and the proposed Electron–Ion Collider in China \cite{Accardi:2012qut, anderle2021electron}. Some of the recent studies in the literature that have focused on gluon Wigner distributions are \cite{Tan:2023vvi,Pasechnik:2023mdd,Hagiwara:2017fye,Boussarie:2018zwg,more2018wigner}. Wigner distributions of gluon are also closely linked to generalized transverse momentum-dependent distributions (GTMDs), which unify and extend the framework of parton distribution functions (PDFs), transverse momentum-dependent distributions (TMDs), and generalized parton distributions (GPDs), offering a broader perspective on the internal dynamics of hadrons \cite{ojha2023quark,maji2022leading,meissner2008generalized,lorce2013structure, kaur2018wigner}.

A particularly interesting scenario arises when considering non-zero skewness, where longitudinal momentum is transferred to the target. This introduces a natural conjugate variable in coordinate space, enabling a new dimension of imaging \cite{brodsky2006hadron, Brodsky:2006ku}. In this context, a boost-invariant longitudinal coordinate, defined as $\sigma=\frac{1}{2}b^-P^+$, plays a pivotal role. The variable $\sigma$ is conjugate to the skewness parameter $\xi$, and allows us to explore the spatial structure of the gluon distribution in the longitudinal direction in a frame-independent way, enriching the phase-space interpretation of GTMDs and Wigner functions.

In this manuscript, we study the Wigner distributions for gluon at non-zero skewness using the dressed quark model.  
This simple yet insightful framework provides an analytically tractable environment to investigate gluon dynamics. We analyze configurations involving transverse polarization, either of the target or the gluon, and observe notable distortions in the distributions. The results for the unpolarized and longitudinal polarization of gluon has been presented in \cite{jana2024gluon}.  These polarization effects are crucial for understanding spin–momentum correlations and are relevant in the context of gluon orbital angular momentum and gluon spin decomposition in the nucleon.

Our analysis reveals a diffraction-like pattern for the gluon Wigner distributions in the $\sigma$ space for configurations involving a transversely polarized gluon and/or a transversely polarized target. This behavior is qualitatively similar to the earlier reported results for unpolarized and longitudinally polarized configurations. This indicate that the gluon Wigner distributions in the boost-invariant longitudinal space are sensitive to both the squared momentum transfer $-t$ and the polarization of the system.

The manuscript is organized as follows. Section 2 introduces the light-front conventions, outlines the relevant kinematic setup, and details the dressed quark model used in our analysis. In Section 3, we define the gluon Wigner distributions with a particular focus on the non-zero skewness regime. Section 3.1 presents the analytical expressions for the gluon Wigner distributions in the boost-invariant longitudinal impact parameter space. Section 4 provides the numerical results and a discussion of the physical implications. Finally, we conclude in Section 5.



\section{Dressed Quark Model and Kinematics}

 We use the dressed quark model to investigate the gluon Wigner distributions. The dressed quark, modeled as a spin-$\frac{1}{2}$ state consisting of a bare quark and a gluon in the leading-order light-front Fock space expansion, offers a simple yet insightful framework for studying gluon Wigner distributions. The model has been used in earlier studies to investigate the other relevant distribution functions, orbital angular momentum, and spin–orbit correlations of quarks and gluons~\cite{mukherjee2015wigner, mukherjee2014quark, ojha2023quark, harindranath1999nonperturbative, harindranath1999orbital}.

We adopt the light-front coordinate system \((x^+, x^-, \boldsymbol{x}_\perp)\), where $x^+$ is the light-front time and $x^-$ is light-front longitudinal spatial coordinate, defined as \(x^\pm = x^0 \pm x^3\)~\cite{harindranath1996introduction}. The total squared momentum transfer to the target is given by \(t = \Delta^2 = (p - p')^2\) and the longitudinal momentum asymmetry between the initial and final target states is characterized by the skewness parameter \(\xi\), defined as
\[
\xi = \frac{\Delta^+}{2P^+},
\]
where \(P^+ = \frac{p^+ + p'^+}{2}\) is the average longitudinal momentum of the target. Choosing a symmetric frame~\cite{brodsky2001light}, in which the average momentum is defined as \(P = \frac{1}{2}(p + p')\), the initial and final target states are parameterized as
\begin{align}
p &= \left((1 + \xi) P^+,\, \frac{\boldsymbol{\Delta}_\perp}{2},\, \frac{m^2 + \boldsymbol{\Delta}_\perp^2/4}{(1 + \xi) P^+} \right), \\
p' &= \left((1 - \xi) P^+,\, -\frac{\boldsymbol{\Delta}_\perp}{2},\, \frac{m^2 + \boldsymbol{\Delta}_\perp^2/4}{(1 - \xi) P^+} \right).
\end{align}
The momentum transfer is then given by
\[
\Delta = p - p' = \left(2\xi P^+,\, \boldsymbol{\Delta}_\perp,\, \frac{t + \boldsymbol{\Delta}_\perp^2}{2\xi P^+} \right),
\]
and the invariant momentum transfer squared is expressed as
\[
t = -\frac{4 \xi^2 m^2 + \boldsymbol{\Delta}_\perp^2}{1 - \xi^2}.
\]

 The dressed quark state with spin \( \sigma \) and momentum \( p \) can be expanded in Fock space up to the two-particle sector, incorporating the quark-gluon configuration \cite{harindranath1999nonperturbative,harindranath1999orbital}
\begin{align}\label{fockse}
\Big|p^+,p_\perp,\sigma  \Big\rangle = &\Phi^\sigma(p) b^\dagger_\sigma(p)
 | 0 \rangle +
 \sum_{\sigma_1 \sigma_2} \int [dp_1]
 \int [dp_2]\nn\\
 &\sqrt{16 \pi^3 p^+}
 \delta^3(p-p_1-p_2) \Phi^\sigma_{\sigma_1 \sigma_2}(p;p_1,p_2)\nn\\
 &b^\dagger_{\sigma_1}(p_1) 
 a^\dagger_{\sigma_2}(p_2)  | 0 \rangle.
\end{align}
The state is expressed as a superposition of a single-quark state and a two-particle quark-gluon state, where the light-front wavefunctions (LFWFs) encapsulate the nonperturbative information. Here, $[dp] =  \frac{dp^{+}d^{2}p_{\perp}}{ \sqrt{16 \pi^3 p^{+}}}$, $\Phi^\sigma(p)$ represents the single-particle wavefunction with momentum $p$ and spin $\sigma$. $\Phi^{\sigma}_{\sigma_1 \sigma_2}(p;p_1,p_2)$ represents the two-particle light-front wavefunction and can be expressed in a boost-invariant form using the relation $\Psi^{\sigma}_{\sigma_1
\sigma_2}(x, q_\perp) =   
\Phi^{\sigma}_{\sigma_1 \sigma_2}
\sqrt{P^+}$.
The momentum variables $(x_i, q_{i \perp})$ are the Jacobi momenta, defined as 
\begin{align}
    p_i^+= x_i p^+, ~~~~~~~~~~q_{i \perp}= k_{i \perp}+x_i p_\perp,
\end{align}
and obey the following conditions
\begin{align}
    \sum_i x_i=1, ~~~~~~~~~\sum_i q_{i\perp}=0.
\end{align}
Assigning $(x_1,q_{1\perp})\equiv (x,q_\perp)$ to the quark, and $(x_2,q_{2\perp})\equiv(x_g,q_{\perp g})$ to the gluon, we obtain
 \begin{align}
x+x_g=&1\Rightarrow x_g=1-x,\\
\mathrm{and} ~~~q_{\perp}+q_{\perp g}=&0 \Rightarrow q_{\perp g}=- q_{\perp}.
\end{align}
The longitudinal momentum fraction of the gluon with respect to the target is $x_g=k^+_g/P^+$, and the corresponding four-momentum of the gluon is 
\begin{align}
    k_g\equiv \Big(x_g P^+, k_{\perp g}, k^-_g \Big)\label{mom_k}.
\end{align}

\section{Gluon Wigner distribution at non-zero skewness}

We investigate the gluon Wigner distributions within the dressed quark model, focusing on scenarios with non-zero skewness. These distributions are constructed using the LFWFs and defined via the Fourier transform of gluon-gluon correlators in longitudinal boost-invariant space.
The general expression for the gluon Wigner distribution is given by~\cite{ji2013probing, more2018wigner, mukherjee2015wigner}:
\begin{align}
     & xW_{\lambda\lambda'}(x, k_\perp, \Delta_\perp, \sigma) = \int \frac{d\xi}{2\pi} e^{i\sigma \cdot \xi} \int \frac{dz^- d^2z_\perp}{2(2\pi)^3 p^+} e^{ik \cdot z} \nn\\
    & \times \left\langle p^+, \frac{\Delta_\perp}{2}, \lambda' \middle| \Gamma^{ij} F^{+i}\left(-\frac{z}{2}\right) \mathcal{W} F^{+j}\left(\frac{z}{2}\right) \mathcal{W}' \middle| p^+, \frac{\Delta_\perp}{2}, \lambda \right\rangle \Bigg|_{z^+ = 0}.
\end{align}
Here, the skewness parameter $\xi$ denotes the longitudinal momentum transfer to the target, and $\sigma$ represents the longitudinal impact parameter space variable conjugate to $\xi$~\cite{maji2022leading}. The field strength tensors $F^{+i}$ are evaluated at two spatially separated points and connected via Wilson lines $\mathcal{W}$ and $\mathcal{W}'$, ensuring gauge invariance. 
The gauge-invariant Wilson line $\mathcal{W}$ connecting two points $z_1$ and $z_2$ is defined as
\begin{align*}    
\mathcal{W}(z_1,z_2)=\mathcal{P}\bigg[exp(-ig\int_{z_1}^{z_2}d\eta^\mu A_\mu(\eta))\Bigg].
\end{align*}
The Wilson line $\mathcal{W}(z_1,z_2)$ has to be evaluated along some path connecting the point $z_1$ and $z_2$. We choose the path such that $z^+=0$ as our fields are defined at $z^+=0$. So, we move in $z^--z_\perp$ plane. For connecting any two general point $(z_1,z_2)$ in $z^--z_\perp$ plane and representing a generic vector $z^\mu=(z^+,z^-,z_\perp)$, we can define the longitudinal gauge link for point separated in longitudinal direction i.e if $z_1=(0,a^-,c_\perp),~ z_2=(0,b^-,c_\perp)$ \cite{bacchetta2016electron},
\begin{align*}    
\mathcal{W}^-(a^-,b^-;c_\perp)=\mathcal{P}\Bigg[exp(-ig\int_{a^-}^{b^-}d\eta^- A^+(0,\eta^-,c_\perp))\Bigg],
\end{align*}
and the transverse gauge link for point separated in transverse direction i.e if $z_1=(0,c^-,a_\perp),~z_2=(0,c^-,b_\perp),$
\begin{align*}    
\mathcal{W}^\perp(a_\perp,b_\perp;c^-)=\mathcal{P}\Bigg[exp(ig\int_{a_\perp}^{b_\perp}d\eta_\perp A_\perp(0,c^-,\eta_\perp))\Bigg].
\end{align*}
Any two points in the $z^--z_\perp$ plane can be connected using the combination of the above-defined longitudinal and transverse gauge links. Since, we are interested in the bi-local operator $F^{+i}\left(-\frac{z}{2}\right) \mathcal{W} F^{+j}\left(\frac{z}{2}\right)$, we choose $z_1=-\frac{z}{2}$ and $z_2=\frac{z}{2}$. We choose to connect these points by a staple-like path of straight lines passing through infinity. So we have 
\begin{align*}
    \mathcal{W}(-\frac{z}{2},\frac{z}{2})&\equiv\mathcal{W}(-\frac{z}{2},\infty)\mathcal{W}(\infty,\frac{z}{2})\\
    \text{where,}~~\mathcal{W}(-\frac{z}{2},\infty)&=\mathcal{W}^-(-\frac{z^-}{2},\infty^-;\frac{z_\perp}{2})\mathcal{W}^\perp(-\frac{z_\perp}{2},\infty_\perp;\infty^-)\\
    \text{and,}~~~~~~\mathcal{W}(\infty,\frac{z}{2})&=\mathcal{W}^\perp(\infty_\perp,\frac{z_\perp}{2};\infty^-)\mathcal{W}^-(\infty^-,\frac{z^-}{2};\frac{z^\perp}{2})
\end{align*}
In light-front gauge we have $A^+=0$, which makes the longitudinal gauge link $(\mathcal{W}^-)$ unity and we are left with only transverse gauge link 
\begin{align*}
     \mathcal{W}(-\frac{z}{2},\frac{z}{2})&=\mathcal{W}^\perp(-\frac{z_\perp}{2},\infty_\perp;\infty^-)\mathcal{W}^\perp(\infty_\perp,\frac{z_\perp}{2};\infty^-)\\
    =&1+ig\int_{-\frac{z_\perp}{2}}^{\frac{z_\perp}{2}}d\eta_\perp \cdot A_\perp(0,\infty^-,\eta_\perp)+\cdot \cdots 
\end{align*}
For our calculations, we retain only the leading-order term (unity), thereby neglecting all higher-order contributions.
As we adopt the light-front gauge $A^+ = 0$, the field strength tensor simplifies to $F^{+i} = \partial^+ A^i$.
The transverse components of the gauge field $A^i$ are expressed in terms of creation and annihilation operators~\cite{harindranath1999orbital}
\begin{align}
    A^{i}\Big(\frac{z}{2}\Big) = &
\sum _{\lambda} \int \frac{dk^{+}d^{2} k_{\perp}}{2k^{+}(2\pi)^3}  
\Big[  \epsilon_{\lambda}^{i}(k) a_{\lambda}(k)e^{-\frac{i}{2}k.z}   +   
\epsilon_{\lambda}^{*i}(k) a^{\dagger}_{\lambda}(k)e^{\frac{i}{2}k.z} \Big].
\end{align}
At twist-two, the gluon Wigner distributions are constructed using the polarization projectors $\Gamma^{ij} \in \{\delta^{ij}_\perp, -i\epsilon^{ij}_\perp, \Gamma^{RR}, \Gamma^{LL} \}$, where $R$ and $L$ denote right- and left-handed polarizations respectively~\cite{lorce2013structure}.

In the dressed quark model, the target state is expanded in terms of Fock components, and we retain only the quark-gluon two-particle sector. The two-particle LFWFs are obtained using light-front Hamiltonian perturbation theory and expressed as \cite{harindranath1999orbital}
\begin{align}
\Psi_{\sigma_1\sigma_2}^{\lambda a}(x,q_\perp) = \frac{g}{\sqrt{2(2\pi)^3}} T^a \frac{\chi_{\sigma_1}^\dagger \, \mathcal{O}(x, q_\perp, m) \, \chi_\lambda \, (\epsilon_{\perp\sigma_2})^* }{D(k_\perp, x)},
\end{align}
where 
{\small \begin{align}
\mathcal{O}(x, q_\perp, m) = 
\frac{1}{\sqrt{1 - x}} 
\left[
-2 \frac{q_\perp}{1 - x} 
- \frac{(\sigma_\perp \cdot q_\perp) \sigma_\perp}{x} 
+ i m \sigma_\perp \frac{(1 - x)}{x}\nn
\right]
\end{align}}
 denotes the spin-momentum structure, and 
\begin{align}
D(k_\perp, x) = m^2 
- \frac{m^2 + k_\perp^2}{x} 
- \frac{k_\perp^2}{1 - x}\nn
\end{align}
is the energy denominator.
Using the LFWFs, the gluon-gluon correlators for different choices of $\Gamma^{ij}$ are expressed as overlaps of wave functions weighted by gluon polarization vectors. The correlators depend on the kinematic variables $x_g = 1 - x$, the skewness $\xi$, transverse momentum $k_{\perp g}$, and transverse momentum transfer to the target $\Delta_\perp$. The polarization vectors $\epsilon^L$ and $\epsilon^R$ are constructed from the Cartesian components via $\epsilon^{L(R)} = \epsilon^1 \mp i \epsilon^2$.

The gluon-gluon correlator functions can be computed using the overlaps of two-particle LFWFs~\cite{more2018wigner}. These correlators depend on the gluon polarization operator \( \Gamma^{ij} \) and the polarization of the target state. Explicitly, the correlators are given as follows:

\noindent
(a) For the unpolarized case, corresponding to \(\Gamma^{ij} = \delta^{ij}_\perp\),
\begin{align}
W^{(\delta^{ij}_\perp)}_{\sigma\sigma'} = -\sum_{\sigma_1, \lambda_1, \lambda_2}
\Psi^{*\,\sigma'}_{\sigma_1 \lambda_1}(x_g', q'_{\perp g})\,
&\Psi^{\sigma}_{\sigma_1 \lambda_2}(y_g, q_{\perp g})\nn\\
&\left(\epsilon^1_{\lambda_2} \epsilon^{*1}_{\lambda_1} + \epsilon^2_{\lambda_2} \epsilon^{*2}_{\lambda_1} \right).
\end{align}
(b) For the longitudinally polarized case with \(\Gamma^{ij} = -i\epsilon^{ij}_\perp\),
\begin{align}
W^{(-i\epsilon^{ij}_\perp)}_{\sigma\sigma'} = -i\sum_{\sigma_1, \lambda_1, \lambda_2}
\Psi^{*\,\sigma'}_{\sigma_1 \lambda_1}(x_g', q'_{\perp g})\,
&\Psi^{\sigma}_{\sigma_1 \lambda_2}(y_g, q_{\perp g})\nn\\
&\left(\epsilon^1_{\lambda_2} \epsilon^{*1}_{\lambda_1} - \epsilon^2_{\lambda_2} \epsilon^{*2}_{\lambda_1} \right).
\end{align}
(c) For the right-handed circular gluon polarization \(\Gamma^{ij} = \Gamma^{RR}\),
\begin{align}
W^{(\Gamma^{RR})}_{\sigma\sigma'} = -\sum_{\sigma_1, \lambda_1, \lambda_2}
\Psi^{*\,\sigma'}_{\sigma_1 \lambda_1}(x_g', q'_{\perp g})\,
\Psi^{\sigma}_{\sigma_1 \lambda_2}(y_g, q_{\perp g})\,
\epsilon^R_{\lambda_2} \epsilon^{*R}_{\lambda_1}. \label{coright}
\end{align}
(d) For left-handed circular gluon polarization \(\Gamma^{ij} = \Gamma^{LL}\)
\begin{align}
W^{(\Gamma^{LL})}_{\sigma\sigma'} = -\sum_{\sigma_1, \lambda_1, \lambda_2}
\Psi^{*\,\sigma'}_{\sigma_1 \lambda_1}(x_g', q'_{\perp g})\,
\Psi^{\sigma}_{\sigma_1 \lambda_2}(y_g, q_{\perp g})\,
\epsilon^L_{\lambda_2} \epsilon^{*L}_{\lambda_1}. \label{coleft}
\end{align}

\noindent
where the kinematic variables are defined as follows: \( x_g = 1 - x \) is the gluon longitudinal momentum fraction, \( x_g' = x_g / (1 - \xi) \), and \( y_g = x_g / (1 + \xi) \). The initial and final momenta of gluon are $(y_g,q_{\perp g})$ and $(x^\prime,q'_{\perp g})$ respectively. The transverse momenta of the gluon in the initial and final states are given by
\begin{align}
q_{\perp g} = -k_\perp + \frac{x_g \Delta_\perp}{2(1 + \xi)}, \quad
q'_{\perp g} = -k_\perp - \frac{x_g \Delta_\perp}{2(1 - \xi)},
\end{align}
and the circular polarization vectors of the gluon are \( \epsilon^{L(R)}_\lambda = \epsilon^1_\lambda \mp i \epsilon^2_\lambda \)~\cite{lorce2013structure}.

The gluon Wigner distributions in boost-invariant longitudinal space are obtained by taking the Fourier transform of correlators with respect to $\xi$:
\begin{align}
\rho^{\Gamma}(x_g, \sigma, \Delta_\perp, k_\perp; S) = \int_0^{\xi_{\text{max}}} \frac{d\xi}{2\pi} e^{i\sigma \cdot \xi} W^{\Gamma}(x_g, \xi, \Delta_\perp, k_\perp; S),
\end{align}
where $S$ denotes the polarization of the target state, and $\xi_{\text{max}}$ is given by
\begin{align}\label{Mandelstem}
\xi_{\text{max}} = \frac{-t}{2m^2} \left(\sqrt{1 + \frac{4m^2}{-t}} - 1 \right), \quad \text{with } -t = \frac{4\xi^2 m^2 + \Delta_\perp^2}{1 - \xi^2}.
\end{align}
For the case of an unpolarized gluon and an unpolarized target, the Wigner distribution is defined as 
\begin{align}
\rho^g_{UU}(x_g, \sigma, \Delta_\perp, k_\perp) &= \frac{1}{2} \Big[ \rho^{\delta^{ij}_\perp}(x, \sigma, \Delta_\perp, k_\perp; +\hat{e}_z) \nn\\
& + \rho^{\delta^{ij}_\perp}(x, \sigma, \Delta_\perp, k_\perp; -\hat{e}_z) \Big].
\end{align}
Substituting the relevant expressions for the gluon-gluon correlator, we obtain an analytic form of the Wigner distribution
\begin{align}\label{wigUU}
    \rho^g_{UU}(x_g,\sigma,t,k_\perp)&=\int_{0}^{\xi_{max}}\frac{d\xi}{4\pi}e^{i\sigma\cdot\xi}~\frac{\alpha_g}{x_g}\Big[-((4(1-\xi^2)k^2_\perp\nn\\
    &-x_g^2\Delta^2_\perp+4\xi x_g k_\perp\cdot\Delta_\perp)((1+x_g^2)+\xi\nn\\
    &(2-3\xi))+4m^2((1-x_g)^2-\xi^2)^2)\Big].
\end{align}
 The integrand includes kinematic and mass-dependent contributions along with a factor
\begin{align}
\alpha_g = \frac{N \sqrt{1 - \xi^2}}{D(q_{\perp g}, y_g) D^*(q'_{\perp g}, x'_g) x_g ((1 - x_g)^2 - \xi^2)^{3/2}}.
\end{align}
Here $N = \frac{g^2 C_F}{2(2\pi)^3}$, with $g$ being the strong coupling constant and $C_F$ is the color factor.


\subsection{Transverse Polarization Case}

The Wigner distributions for transversely polarized gluons, in conjunction with either an unpolarized or longitudinally polarized target, can be expressed in terms of helicity-dependent correlators~\cite{more2018wigner}. For an unpolarized target, the gluon Wigner distributions with right- and left-handed transverse polarizations are defined as the helicity averages:
\begin{align}
    \rho^{gR}_{UT}(x_g,\sigma,\Delta_\perp,k_\perp) &= \frac{1}{2}\Big[\rho^{\Gamma^{RR}}(x_g,\sigma,\Delta_\perp,k_\perp;+\hat{e}_z) \nn\\ 
    &+ \rho^{\Gamma^{RR}}(x_g,\sigma,\Delta_\perp,k_\perp;-\hat{e}_z)\Big], \\
    \rho^{gL}_{UT}(x_g,\sigma,\Delta_\perp,k_\perp) &= \frac{1}{2}\Big[\rho^{\Gamma^{LL}}(x_g,\sigma,\Delta_\perp,k_\perp;+\hat{e}_z) \nn\\
    & + \rho^{\Gamma^{LL}}(x_g,\sigma,\Delta_\perp,k_\perp;-\hat{e}_z)\Big].
\end{align}
For a longitudinally polarized target, the corresponding helicity-difference Wigner distributions are given by:
\begin{align}
    \rho^{gR}_{LT}(x_g,\sigma,\Delta_\perp,k_\perp) &= \frac{1}{2}\Big[\rho^{\Gamma^{RR}}(x_g,\sigma,\Delta_\perp,k_\perp;+\hat{e}_z) \nn\\ &- \rho^{\Gamma^{RR}}(x_g,\sigma,\Delta_\perp,k_\perp;-\hat{e}_z)\Big], \\
    \rho^{gL}_{LT}(x_g,\sigma,\Delta_\perp,k_\perp) &= \frac{1}{2}\Big[\rho^{\Gamma^{LL}}(x_g,\sigma,\Delta_\perp,k_\perp;+\hat{e}_z) \nn\\
    &- \rho^{\Gamma^{LL}}(x_g,\sigma,\Delta_\perp,k_\perp;-\hat{e}_z)\Big].
\end{align}
In the case of a transversely polarized target, the gluon Wigner distributions depend on the polarization of the gluon and are defined as:
\begin{align}
    \rho_{TT}^{g(R)i}(x_g,\sigma,\Delta_\perp,k_\perp) &= \frac{1}{2}\Big[\rho^{\Gamma^{RR}}(x_g,\sigma,\Delta_\perp,k_\perp;+\hat{e}_i) \nn\\
    &+ \rho^{\Gamma^{RR}}(x_g,\sigma,\Delta_\perp,k_\perp;-\hat{e}_i)\Big], \\
    \rho_{TT}^{g(L)i}(x_g,\sigma,\Delta_\perp,k_\perp) &= \frac{1}{2}\Big[\rho^{\Gamma^{LL}}(x_g,\sigma,\Delta_\perp,k_\perp;+\hat{e}_i) \nn\\ 
    & + \rho^{\Gamma^{LL}}(x_g,\sigma,\Delta_\perp,k_\perp;-\hat{e}_i)\Big], \\
    \rho^{gi}_{TU}(x_g,\sigma,\Delta_\perp,k_\perp) &= \frac{1}{2}\Big[\rho^{\delta_\perp^{ij}}(x_g,\sigma,\Delta_\perp,k_\perp;+\hat{e}_i) \nn\\
    &+ \rho^{\delta_\perp^{ij}}(x_g,\sigma,\Delta_\perp,k_\perp;-\hat{e}_i)\Big], \\
    \rho^{gi}_{TL}(x_g,\sigma,\Delta_\perp,k_\perp) &= \frac{1}{2}\Big[\rho^{-i\epsilon_\perp^{ij}}(x_g,\sigma,\Delta_\perp,k_\perp;+\hat{e}_i)\nn\\
    &+ \rho^{-i\epsilon_\perp^{ij}}(x_g,\sigma,\Delta_\perp,k_\perp;-\hat{e}_i)\Big].
\end{align}
Here, \( \hat{e}_i \) denotes the transverse spin direction of the target, where \( i = x, y \). The transverse polarization states are defined in terms of helicity eigenstates, for example,
\[
|\pm \hat{e}_x \rangle = \frac{1}{\sqrt{2}}\left( \left| \tfrac{1}{2} \right\rangle \pm \left| -\tfrac{1}{2} \right\rangle \right).
\]

The analytic expressions for the gluon Wigner distributions in the boost-invariant longitudinal position space ($\sigma$) corresponding to various combinations of gluon and target polarizations are derived using two-particle light-front wave functions. The explicit results for the transverse polarization configurations are presented in Eqs.~(\ref{Wig1})–(\ref{Wig2}). These include the cases of transversely polarized gluons with unpolarized and longitudinally polarized targets, as well as transversely polarized targets. The integrals involve the skewness parameter $\xi$, and the resulting Wigner distributions depend on the set of kinematic variables $(x_g, \xi, k_\perp^2, \Delta_\perp^2, k_\perp \cdot \Delta_\perp)$. The dependence on the transverse momentum transfer $\Delta_\perp$ is re-expressed in terms of the Mandelstam variable $t$, using the relation in Eq.~(\ref{Mandelstem}), prior to performing the Fourier transform to the $\sigma$-space. The analytical expressions for $\rho_{UU}^g$, $\rho_{TU}^{gx}$, and $\rho_{TL}^{gx}$ in the dressed quark model are mentioned in the Ref. \cite{more2018wigner} for zero skewness. The expression for $\rho_{UU}^g$, $\rho_{TU}^{gx}$, and $\rho_{TL}^{gx}$ for non zero skewness listed here in Eqs.(\ref{wigUU}), (\ref{wigTU}), and (\ref{Wig2}) reproduces the result of \cite{more2018wigner} in the limit $\xi\rightarrow 0$. 
\begin{align}
    \rho^{gR}_{UT}(x_g,\sigma,t,k_\perp)=&\int_{0}^{\xi_{max}}\frac{d\xi}{4\pi}e^{i\sigma\cdot\xi}~\frac{\alpha_g}{x_g}\Big[-4m^2((1-x_g)^2\nn\\
    &-\xi^2)^2+(x_g^2+1-\xi^2)(-4(1-\xi^2)k^2_\perp\nn\\
    &+x_g^2\Delta^2_\perp-4x_g(\xi k_\perp\cdot\Delta_\perp-i(k_1\Delta_2\nn\\
    &-k_2\Delta_1)))\Big],\label{Wig1}\\
    \rho^{gL}_{UT}(x_g,\sigma,t,k_\perp)=&\int_{0}^{\xi_{max}}\frac{d\xi}{4\pi}e^{i\sigma\cdot\xi}~\frac{\alpha_g}{x_g}\Big[-4m^2((1-x_g)^2\nn\\
       &-\xi^2)^2-(1+x_g^2-\xi^2)(4(1-\xi^2)k_\perp^2\nn\\
       &-x_g^2\Delta_\perp^2+4x_g\xi k_\perp\cdot\Delta_\perp+4ix_g(k_1\Delta_2\nn\\
       &-k_2\Delta_1))\Big],\\
      \rho^{gR}_{LT}(x_g,\sigma,t,k_\perp)=&\int_{0}^{\xi_{max}}\frac{d\xi}{4\pi}e^{i\sigma\cdot\xi}~\frac{\alpha_g}{x_g}\Big[4m^2((1-x_g)^2-\xi^2)^2\nn\\
      &+(x_g^2-1+\xi^2)(-4(1-\xi^2)k^2_\perp+x_g^2\Delta^2_\perp\nn\\
      &-4x_g(\xi k_\perp\cdot\Delta_\perp-i(k_1\Delta_2-k_2\Delta_1)))\Big],\\
       \rho^{gL}_{LT}(x_g,\sigma,t,k_\perp)=&\int_{0}^{\xi_{max}}\frac{d\xi}{4\pi}e^{i\sigma\cdot\xi}~\frac{\alpha_g}{x_g}\Big[-4m^2((1-x_g)^2\nn\\
         &-\xi^2)^2-(1-x_g^2+\xi^2)(4(1-\xi^2)k_\perp^2\nn\\
         &-x_g^2\Delta_\perp^2+4x_g\xi k_\perp\cdot\Delta_\perp+4ix_g(k_1\Delta_2\nn\\
         &-k_2\Delta_1))\Big],\\
         \rho_{TT}^{g(R)x}(x_g,\sigma,t,k_\perp)=&\int_0^{\xi_{max}}\frac{d\xi}{4\pi}e^{i\sigma\cdot\xi}~\alpha_g\Big[4im(((1-x_g)^2+\xi^2)\nn\\
    &(2\xi k_1+2ik_2-x_g\Delta_1)-2(1-x_g)\xi(2k_1\nn\\
    &+2i\xi k_2-ix_g\Delta_2))\Big],\\
   \rho_{TT}^{g(L)x}(x_g,\sigma,t,k_\perp)=&\int_0^{\xi_{max}}\frac{d\xi}{4\pi}e^{i\sigma\cdot\xi}~\alpha_g\Big[4m(((1-x_g)^2+\xi^2)\nn\\
     &(2(i\xi k_1+k_2)-ix_g\Delta_1)-2(1-x_g)\nn\\
     &\xi(2(ik_1+\xi k_2)-x_g\Delta_2))\Big],
     \end{align}
    \begin{align}
\rho_{TU}^{gx}(x_g,\sigma,t,k_\perp)=&\int_0^{\xi_{max}}\frac{d\xi}{4\pi}e^{i\sigma\cdot\xi}~\alpha_g\Big[4im(2\xi(x_g^2-1\nn\\
&+\xi^2)k_1-x_g((1-x_g)^2+\xi^2)\Delta_1)\Big],\label{wigTU}\\
\rho_{TL}^{gx}(x_g,\sigma,t,k_\perp)=&\int_0^{\xi_{max}}\frac{d\xi}{4\pi}e^{i\sigma\cdot\xi}~\alpha_g\Big[8m(((1-x_g)^2+\xi^2\nn\\
     &-2(1-x_g)\xi^2)k_2+(1-x_g)x_g\xi\Delta_2)\Big].\label{Wig2}
\end{align}
\begin{figure}[htp!]
\begin{minipage}[c]{1\textwidth}
\small{}\includegraphics[width=8.6cm,height=5.2cm,clip]{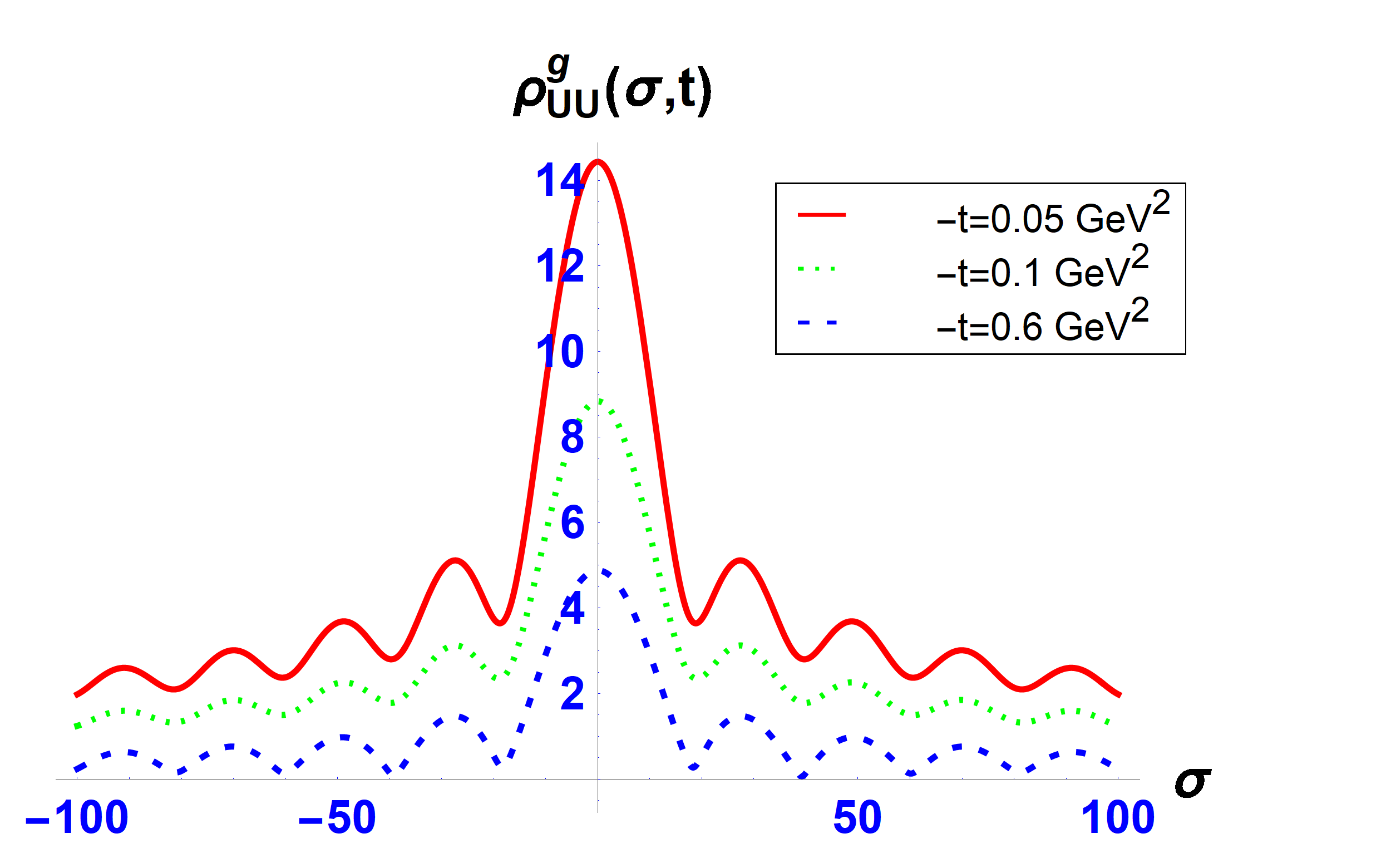}\\
\end{minipage}
\caption{\label{FigWiguu}The first moment of the gluon Wigner distribution $\rho^g_{UU}$ for various values of $-t$, corresponding to the unpolarized gluon in an unpolarized target.}
\end{figure}
%
%
\section{Numerical Analysis and Discussion}
%
The Wigner distributions are plotted for three different values of the squared momentum transfer, $-t = 0.05$, $0.1$, and $0.6~\text{GeV}^2$, in order to study the impact of $-t$ on the transversely polarized gluon Wigner distributions in $\sigma$ space. For all cases, the transverse momentum transfer $\Delta_\perp$ is chosen to be perpendicular to the gluon transverse momentum $k_\perp$, thereby suppressing contributions from terms proportional to $k_\perp \cdot \Delta_\perp$. The mass of the dressed quark is taken to be $0.0033~\text{GeV}$\\
Figure 1 displays the Wigner distribution for an unpolarized gluon in an unpolarized target, presented in the longitudinal impact parameter space. Figure 2 shows the Wigner distributions in $\sigma$ space for cases where the gluon, the target, or both are transversely polarized. In all the plots of Figs. 1 and 2, the gluon longitudinal momentum fraction is fixed at $x_g = 0.7$, and the transverse momentum is chosen as $k_\perp = 0.2\hat{x}$ GeV. \\
Figure 2 illustrates the distortions in the gluon Wigner distributions induced by the transverse polarization of the gluon and the target state. Figs. 2(a) and 2(b) show the modification of the distribution when the gluon is transversely polarized in unpolarized  and longitudinally polarized target. Figs. 2(c) and 2(d) show the plots when the target itself is transversely polarized. The magnitude of the gluon Wigner distributions is significantly reduced when the target is transversely polarized compared to the cases of longitudinal polarization or an unpolarized target. For the distributions $\rho_{UT}$ and $\rho_{LT}$, the peak magnitude increases monotonically as $-t$ decreases. In contrast, for $\rho_{TT}$ and $\rho_{TL}$, the peak height exhibits a non-monotonic dependence on $-t$, reaching a maximum at $-t = 0.1~\text{GeV}^2$ among the three chosen values of $-t$.
\begin{figure}[htp!]
\begin{minipage}[c]{1\textwidth}
\small{(a)}\includegraphics[width=8cm,height=4.6cm,clip]{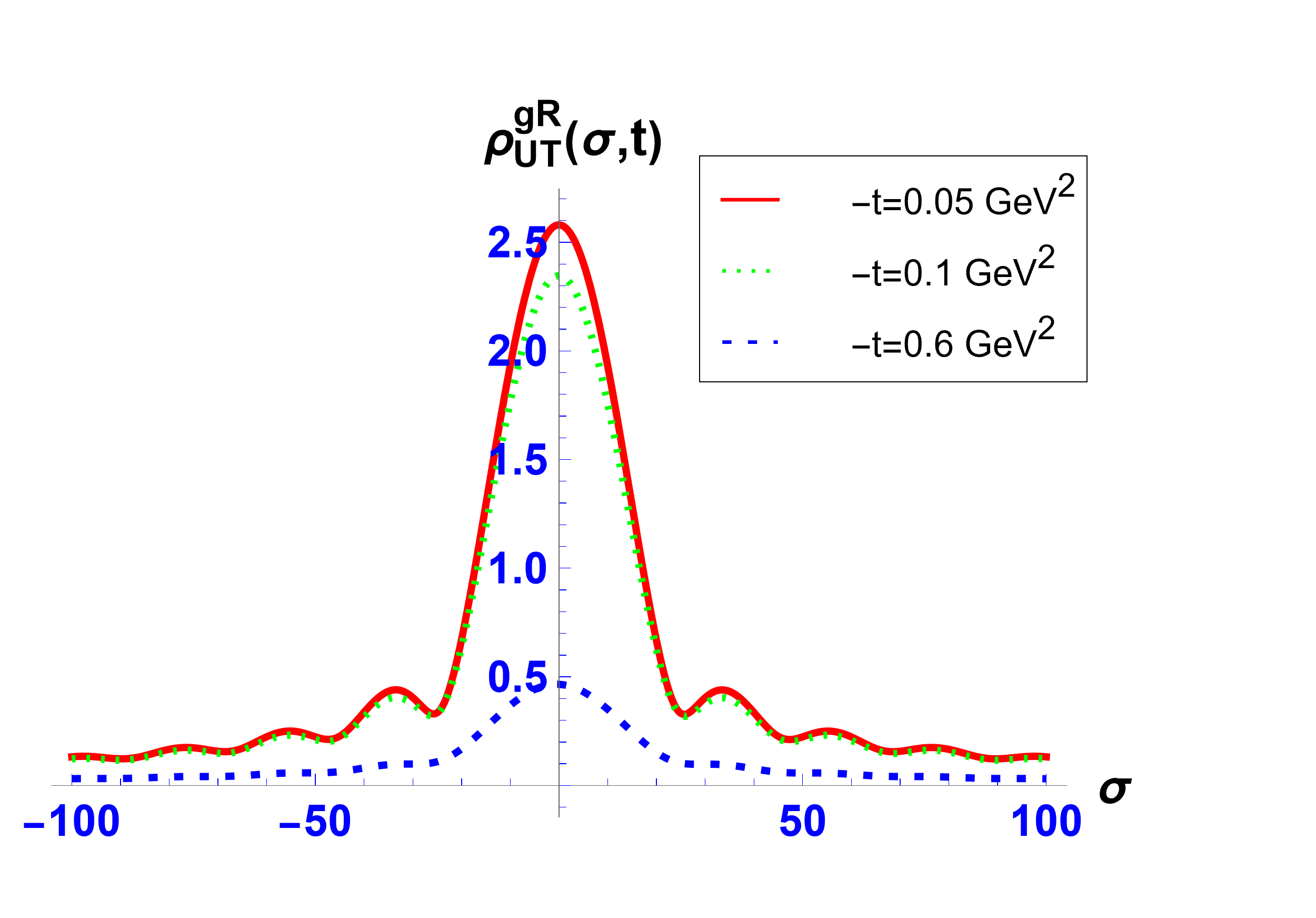}\\
\small{(b)}\includegraphics[width=8cm,height=4.6cm,clip]{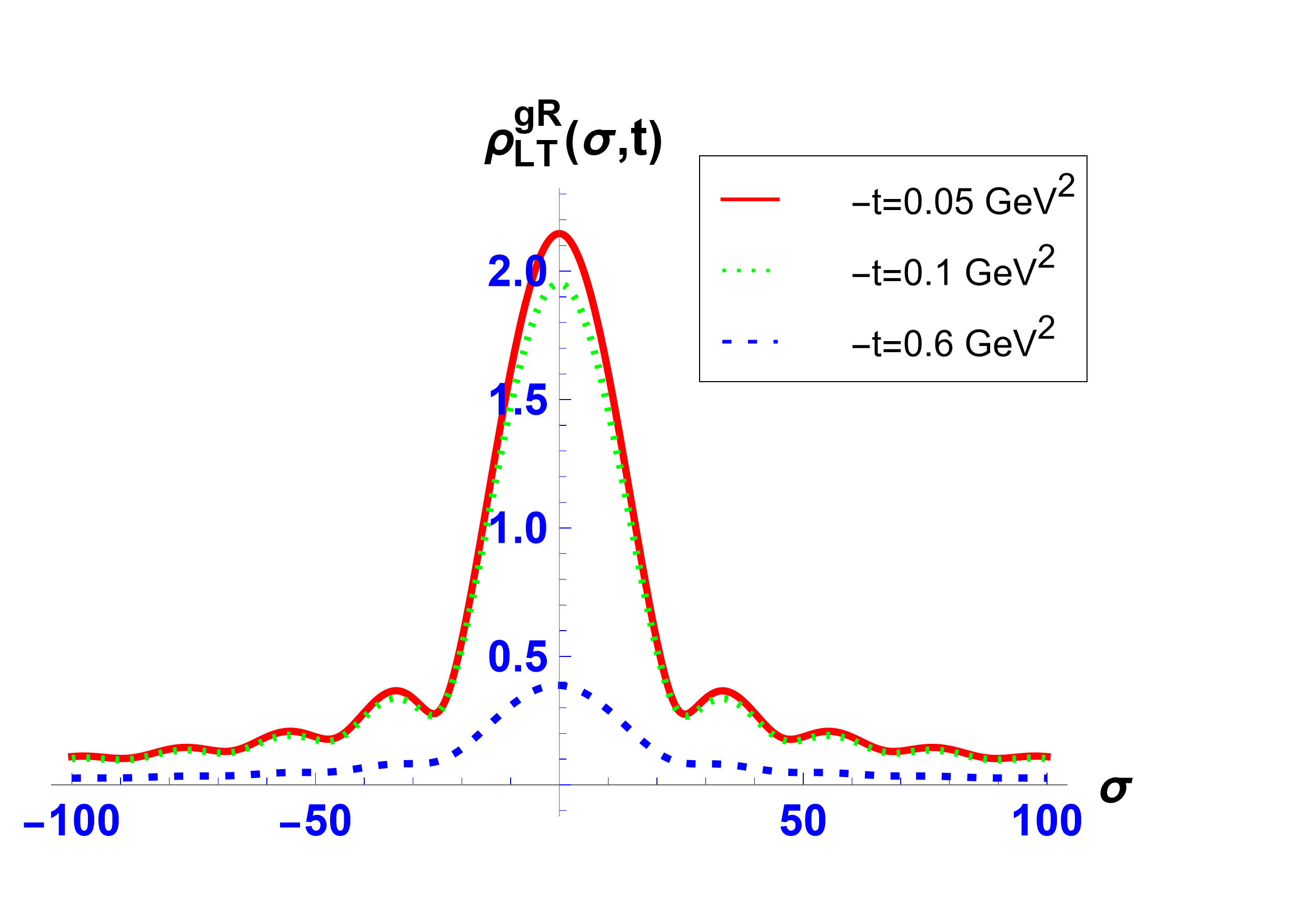}\\
\small{(c)}\includegraphics[width=8cm,height=4.6cm,clip]{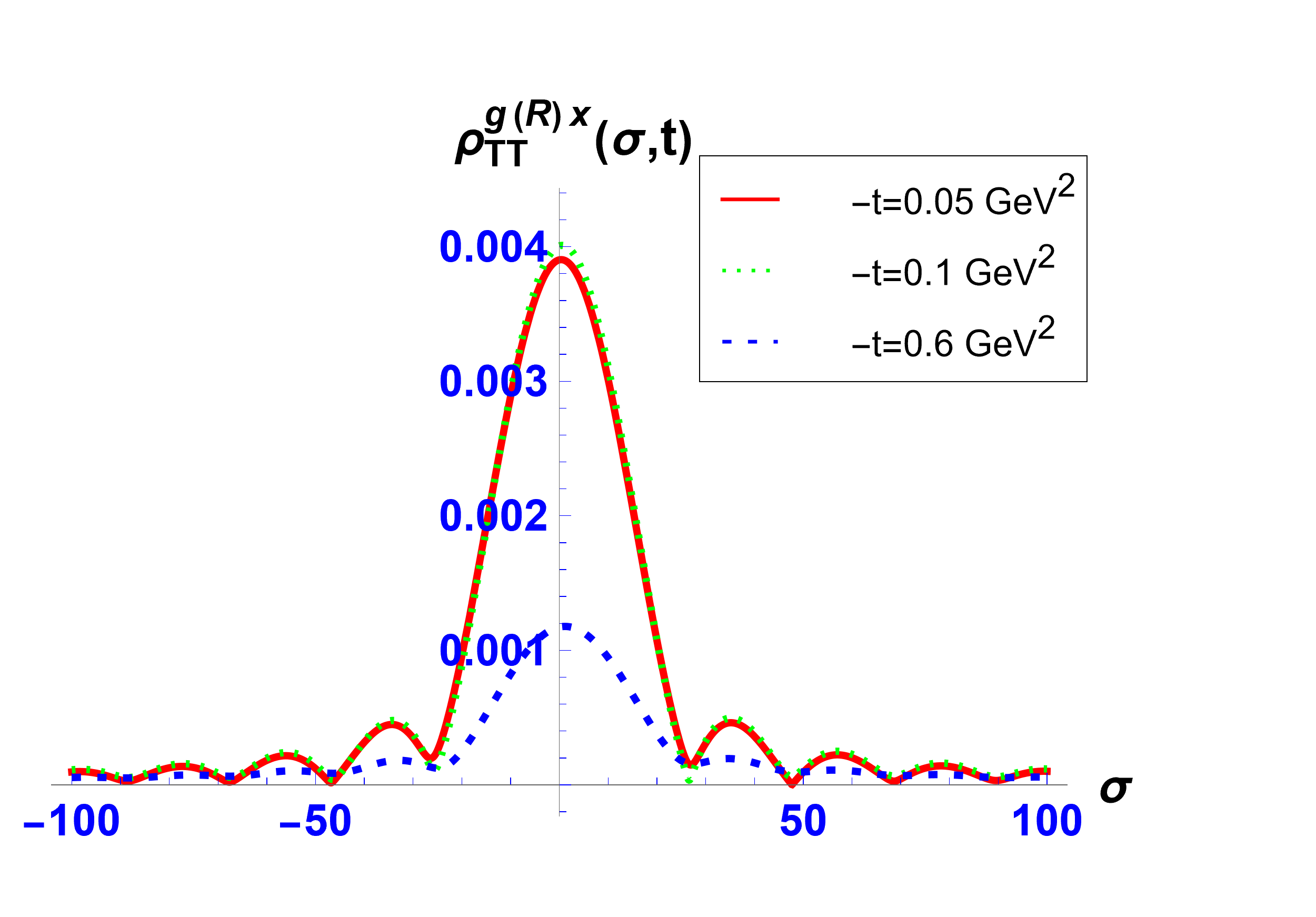}\\
\small{(d)}\includegraphics[width=8cm,height=4.6cm,clip]{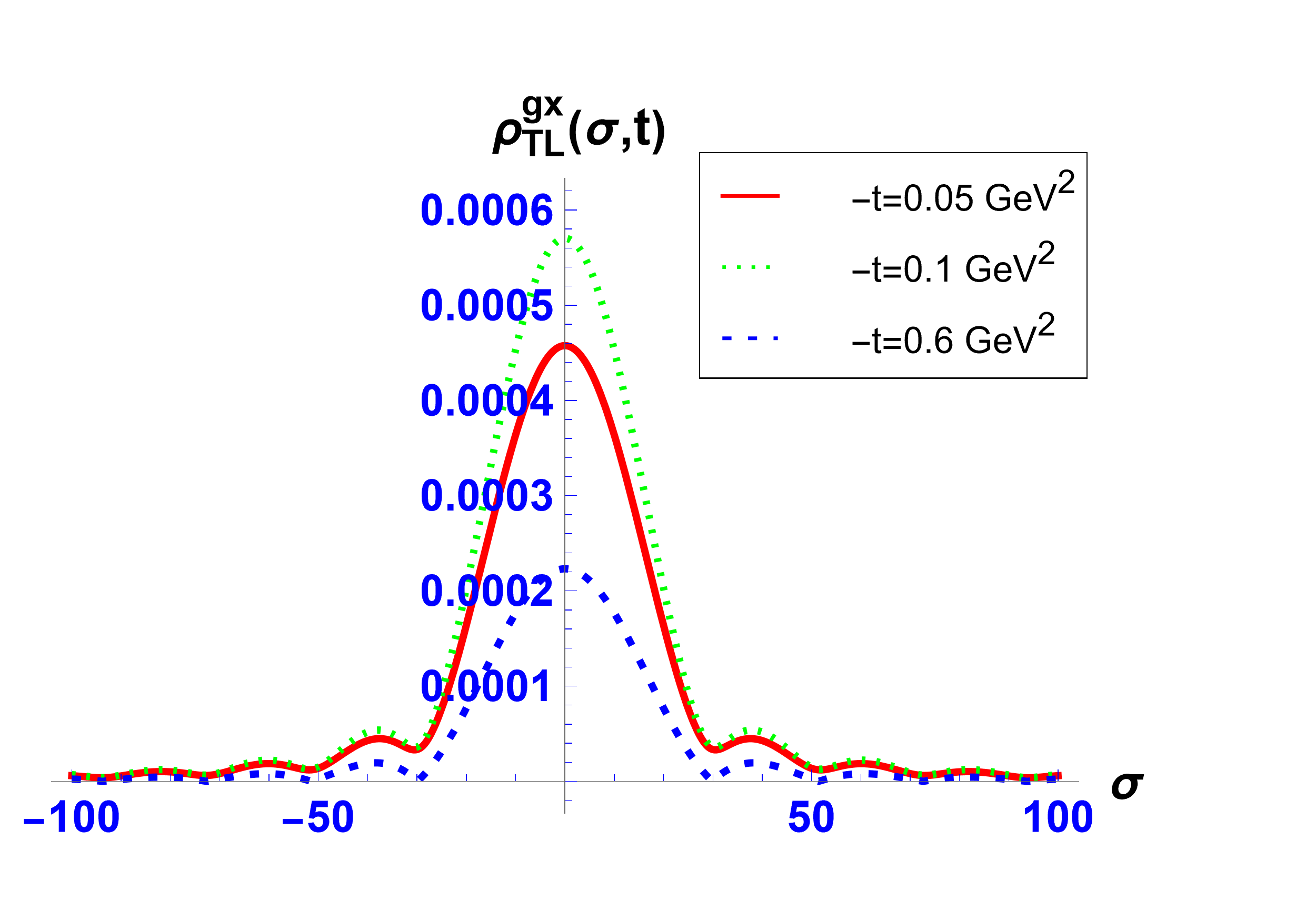}\\
\end{minipage}
\caption{\label{FigWig} The first moment of the gluon Wigner distribution for a) transversely polarized gluon in unplolarized target $\rho^{gR}_{UT}$ b) transversely polarized gluon in longitudinally polarized target $\rho^{gR}_{LT}$ c) $\rho^{g(R)x}_{TT}$ both the gluon and the target are transversely polarized d) $\rho^{gx}_{TL}$ longitudinally polarized gluon and transversely polarized target}
\end{figure}
A common feature across all the distributions is the emergence of a diffraction-like pattern in $\sigma$ space. The results indicate that such diffraction-like patterns persist in the longitudinal impact parameter space even when either the gluon or the target is transversely polarized, with the peak of the distribution exhibiting sensitivity to the magnitude of $-t$. Similar diffraction patterns have also been reported in earlier studies for gluons in cases involving unpolarized and longitudinally polarized gluons and targets within the same model \cite{jana2024gluon}.

\section{Conclusion}
We investigated the gluon Wigner distributions for transverse polarization in the dressed quark model incorporating non-zero skewness, within the light-front Hamiltonian formalism. Using the two-particle LFWFs, we evaluated the gluon-gluon correlators for different transverse polarization configurations and derived the associated Wigner distributions in the longitudinal boost-invariant impact parameter space ($\sigma$). We obtained the analytical expressions for the gluon Wigner distributions, for the cases when either the gluon or the target or both are transversely polarized. \\ 
The gluon Wigner distributions $\rho^g_{UU}$, $\rho^{gR}_{UT}$, $\rho^{gR}_{LT}$, $\rho^{g(R)x}_{TT}$, and $\rho^{gx}_{TL}$ are analyzed in $\sigma$ space through the two-dimensional plots shown in Figs. 1 and 2. All distributions exhibit a diffraction-like oscillatory behavior and sensitivity to the momentum transfer $-t$, consistent with earlier findings for gluon Wigner distributions in unpolarized and longitudinally polarized configurations reported in \cite{jana2024gluon}. Similar oscillatory diffraction patterns have also been observed in studies of quark Wigner distributions within the dressed quark model and the light-front quark-diquark  model \cite{ojha2023quark,maji2022leading}. While quark Wigner distributions with transverse polarization have been explored in various models, to the best of our knowledge, the present work provides the first analysis of gluon Wigner distributions with transverse polarization in the boost-invariant longitudinal impact parameter space. 
\bibliographystyle{elsarticle-num}
\bibliography{Ref.bib}
\end{document}